\documentclass[12pt]{article}
\usepackage{bm,amsmath,amssymb,graphicx}

\begin{document}
\pagenumbering{arabic}
\begin{titlepage}

\title{Massive conformal gravity}

\author{F. F. Faria$\,^{*}$ \\
Centro de Ci\^encias da Natureza, \\
Universidade Estadual do Piau\'i, \\ 
64002-150 Teresina, PI, Brazil}

\date{}
\maketitle

\begin{abstract}
In this article we construct a massive theory of gravity that is invariant 
under conformal transformations. The massive action of the theory depend on 
the metric tensor and a scalar field, which are considered as the only field 
variables. We find the vacuum field equations of the theory and analyze its 
weak-field approximation and Newtonian limit.
\end{abstract}

\thispagestyle{empty}
\vfill
 \par
\bigskip
\noindent * fff@uespi.br \par
\end{titlepage}
\newpage


\section{Introduction}


The study of massive gravity started with Fierz and Pauli \cite{Fierz}, who  
constructed an action describing a free massive spin 2 particle in flat 
spacetime. It was realized later that the Fierz-Pauli theory coupled to a 
source is different from linearized general relativity in the massless limit 
\cite{vDVZ}. This is known as the van Dam-Veltman-Zakharov (vDVZ) 
discontinuity. In order to cure this discontinuity, Vainshtein proposed 
adding nonlinear effects to the Fierz-Pauli theory \cite{Vainshtein}. 
The Vainshtein theory, however, has an extra degree of freedom known as the 
Boulware-Deser (BD) ghost \cite{Boulware}. Some nonlinear massive gravity 
theories developed recently \cite{Rham} eliminate the BD ghost but give 
rise to unstable cosmological solutions \cite{Felice}. The solution of this 
problem leads to massive gravity theories where the Lorentz invariance is 
broken \cite{Comelli}.

It is well know that the theories of elementary particles are invariant under 
Lorentz transformations. In addition, these theories present local conformal 
symmetry. Similarly, it is reasonable to expect that the gravity theory be 
invariant under coordinate transformations and conformal transformations. 
The usual procedures to obtain a conformally invariant gravity theory are 
either to adopt the Weyl action \cite{Weyl} or the Einstein-Hilbert action 
conformally coupled to a scalar field \cite{Dirac}. Several works based on 
the Weyl action have been carried out in the literature (see, for example, 
refs. \cite{Mannheim1, Kazanas, Mannheim2}).       

In this article we address a conformally invariant massive gravity theory 
based on both the Weyl action and the Einstein-Hilbert action conformally 
coupled to a scalar field. In section 2 we construct the massive conformal 
gravitational action and derive the vacuum field equations of the theory. In 
section 3 we investigate the limit of the theory in which the fields are weak. 
In section 4 we find the Newtonian limit of the theory. Finally, in section 5 
we present our conclusions.


\section{Massive gravity with conformal invariance}


A conformal transformation is a change of the spacetime geometry that alters 
the length scales. The conformal transformation of the spacetime metric is 
defined by
\begin{equation}
\tilde{g}_{\mu\nu}=e^{2\theta(x)}\,g_{\mu\nu},
\label{1} 
\end{equation}
where $\theta(x)$ is an arbitrary function of the spacetime coordinates. 
With the help of eq. (1), it is possible to verify that the Weyl tensor
\begin{eqnarray}
C^{\alpha}\,\!\!_{\mu\beta\nu} &=& R^{\alpha}\,\!\!_{\mu\beta\nu} 
+ \frac{1}{2}\left( \delta^{\alpha}\,\!\!_{\nu}R_{\mu\beta} 
- \delta^{\alpha}\,\!\!_{\beta}R_{\mu\nu} + g_{\mu\beta}R^{\alpha}\,\!\!_{\nu} 
- g_{\mu\nu}R^{\alpha}\,\!\!_{\beta} \right) \nonumber \\ && 
+\frac{1}{6}\left(  \delta^{\alpha}\,\!\!_{\beta}g_{\mu\nu} 
- \delta^{\alpha}\,\!\!_{\nu}g_{\mu\beta}\right)R
\label{2}
\end{eqnarray}
is conformally invariant, where $R^{\alpha}\,\!\!_{\mu\beta\nu}$ is the 
Riemann tensor, $R_{\mu\nu} = R^{\alpha}\,\!\!_{\mu\alpha\nu}$ is the Ricci 
tensor and $R = g^{\mu\nu}R_{\mu\nu}$ is the scalar curvature.   

The square of the Weyl tensor leads to the unique gravitational action 
constructed out of the metric tensor only that is invariant under conformal 
transformations. It is given by  
\begin{equation}
S_{g} = - \frac{\alpha}{2kc} \int{d^{4}x} \, \sqrt{-g}\left(C^{\alpha\beta
\mu\nu}C_{\alpha\beta\mu\nu} \right),
\label{3}
\end{equation}
where $\alpha$ is a dimensionless constant and $k = 16\pi G/c^4$ ($G$ is the 
gravitational constant and $c$ is the speed of light in vacuum). The Weyl 
action (3) is of fourth order with respect to the metric derivatives. However, 
by introducing a scalar field it is possible to construct conformally invariant 
gravitational actions having at most second order derivatives of the metric. 
The simplest of such actions reads 
\begin{equation}
S_{g} = \frac{\beta}{2kc} \int{d^{4}x} \, \sqrt{-g}\left( \varphi^{2}R 
+ 6\partial_{\mu}\varphi\partial^{\mu}\varphi \right),
\label{4}
\end{equation}
where $\beta$ is a dimensionless constant and $\varphi$ is a scalar field that 
transforms as 
\begin{equation}
\tilde{\varphi}=e^{-\theta} \varphi
\label{5}
\end{equation}
under a conformal transformation. Note that the action (4) is invariant under 
conformal transformations after the appropriate integration of the boundary 
term \cite{Fujii}. 

The actions (3) and (4) are the main candidates to form a massive 
gravitational action with conformal symmetry. In analogy with other massive 
theories, it is expected that the mass term of a massive gravitational 
action be of lower order with respect to the metric derivatives than the 
massless term of the action. Thus a natural choice of a conformally 
invariant massive gravitational action is given by   
\begin{equation}
S_{g} = - \frac{1}{2kc} \int{d^{4}x} \, \sqrt{-g}\left[ \alpha 
C^{\alpha\beta\mu\nu}C_{\alpha\beta\mu\nu} - \beta \lambda^{-2}\left(\varphi^{2}R 
+ 6 \partial_{\mu}\varphi\partial^{\mu}\varphi \right) \right],
\label{6}
\end{equation}
where $\lambda = \hbar/mc$ ($\hbar$ is the Planck constant and $m$ is the graviton 
mass).

Varying the action (6) with respect to $g^{\mu\nu}$ and $\varphi$ in vacuum, 
we obtain the field equations
\begin{equation}
2 \alpha W_{\mu\nu} - \beta \lambda^{-2} \left[\varphi^{2}G_{\mu\nu} +  
6 \partial_{\mu}\varphi\partial_{\nu}\varphi - 3g_{\mu\nu}\partial_{\rho}\varphi
\partial^{\rho}\varphi + g_{\mu\nu} \Box \varphi^{2} - \nabla_{\mu}\nabla_{\nu} 
\varphi^{2} \right] = 0,
\label{7}
\end{equation}
and
\begin{equation}
\Box \varphi - \frac{1}{6} R \varphi = 0,
\label{8}
\end{equation}
respectively, where
\begin{equation}
W_{\mu\nu} = \nabla^{\alpha}\nabla^{\beta}C_{\mu\alpha\nu\beta} 
-\frac{1}{2} R^{\alpha\beta}C_{\mu\alpha\nu\beta} 
\label{9}
\end{equation}
is the Bach tensor,
\begin{equation}
G_{\mu\nu} = R_{\mu\nu} - \frac{1}{2}g_{\mu\nu}R
\label{10}
\end{equation}
is the Einstein tensor, and
\begin{equation}
\Box \varphi = \nabla_{\rho}\nabla^{\rho} \varphi = 
\frac{1}{\sqrt{-g}}\partial_{\rho}\left( \sqrt{-g} \partial^{\rho}
\varphi \right)
\label{11}
\end{equation} 
is the generally covariant d'Alembertian for a scalar field. The field 
equations (7) and (8) are rather intricate, and it is not easy to find any 
simple solution of these equations. However, the Newtonian limit of the 
theory yields a simple and interesting solution, as we shall see in section 4.


\section{The weak-field approximation}


By imposing the weak-field approximations 
\begin{equation}
g_{\mu\nu} = g^{(0)}_{\mu\nu} + h_{\mu\nu} = \eta_{\mu\nu} + h_{\mu\nu}, 
\label{12}
\end{equation}
\begin{equation}
\varphi = \varphi^{(0)}(1 + \sigma)  = \sqrt{\frac{2\alpha}{\beta}}(1 + \sigma),
\label{13}
\end{equation}
to eqs. (7) and (8), and neglecting terms of second order in $h_{\mu\nu}$ and 
$\sigma$, we obtain the linearized field equations
\begin{eqnarray}
&& \partial_{\rho}\partial^{\rho}\bar{R}_{\mu\nu} 
- \frac{1}{3}\partial_{\mu}\partial_{\nu}\bar{R} 
- \frac{1}{6}\eta_{\mu\nu}\partial_{\rho}\partial^{\rho}\bar{R} 
\nonumber \\ && 
- \lambda^{-2} \left( \bar{R}_{\mu\nu} - \frac{1}{2}\eta_{\mu\nu}
\bar{R} -2\partial_{\mu}\partial_{\nu}\sigma
+ 2 \eta_{\mu\nu}\partial_{\rho}\partial^{\rho}\sigma \right)= 0,
\label{14}
\end{eqnarray}
and
\begin{equation}
\partial_{\rho}\partial^{\rho}\sigma - \frac{1}{6}\bar{R} = 0,
\label{15}
\end{equation}
respectively, where
\begin{equation}
\bar{R}_{\mu\nu} = \frac{1}{2} \left( \partial_{\mu}\partial^{\sigma}
h_{\sigma\nu} + \partial_{\nu}\partial^{\sigma}h_{\sigma\mu} 
- \partial_{\sigma}\partial^{\sigma}
h_{\mu\nu} - \partial_{\mu}\partial_{\nu}h  \right)
\label{16}
\end{equation}
is the linearized Ricci tensor, and
\begin{equation}
\bar{R} =  \partial^{\mu}\partial^{\nu}h_{\mu\nu} 
- \partial_{\rho}\partial^{\rho}h 
\label{17}
\end{equation} 
is the linearized scalar curvature, with $h = h^{\rho}\,\!\!_{\rho} 
= \eta^{\mu\nu}h_{\mu\nu}$.

The linearized field equations (14) and (15) are invariant under the 
coordinate gauge transformation
\begin{equation}
h_{\mu\nu} \rightarrow h_{\mu\nu} + \partial_{\mu}\xi_{\nu} + 
\partial_{\nu}\xi_{\mu},
\label{18}
\end{equation}
where $\xi^{\mu}$ is an arbitrary spacetime dependent vector field, and
under the conformal gauge transformations
\begin{equation}
h_{\mu\nu} \rightarrow h_{\mu\nu} + \eta_{\mu\nu}\Lambda,
\label{19}
\end{equation}
\begin{equation}
\sigma \rightarrow \sigma - \frac{1}{2}\Lambda,
\label{20}
\end{equation}
where $\Lambda$ is an arbitrary spacetime dependent scalar field.

We may impose the coordinate gauge condition
\begin{equation}
\partial^{\mu}h_{\mu\nu} - \frac{1}{2}\partial_{\nu}h = 0,
\label{21}
\end{equation}
which fixes the coordinate gauge freedom up to a residual coordinate gauge 
parameter satisfying $\partial_{\rho}\partial^{\rho}\xi_{\mu} = 0$, and the 
conformal gauge condition \footnote{We can instead impose the unitary gauge 
$\sigma = 0$. These two gauge conditions give the same classical results, 
as we shall see in the next section. However, the unitary gauge is not 
suitable for a quantum analysis \cite{Hooft}, since it breaks the conformal 
symmetry.}
\begin{equation}
\partial^{\mu}\partial^{\nu}h_{\mu\nu} 
- \partial_{\rho}\partial^{\rho}h  - 6\lambda^{-2}\sigma =  0,
\label{22}
\end{equation}
which fixes the conformal gauge freedom up to a residual conformal gauge 
parameter satisfying $(\partial_{\rho}\partial^{\rho} - \lambda^{-2})
\Lambda = 0$. Combining eqs. (14), (15), (21) and (22), we arrive at
\begin{equation}
\left(\partial_{\rho}\partial^{\rho} - \lambda^{-2} \right)\partial_{\sigma}
\partial^{\sigma}h_{\mu\nu}  = 0,
\label{23}
\end{equation}
\begin{equation}
\left(\partial_{\rho}\partial^{\rho} - \lambda^{-2} \right)\sigma = 0.
\label{24}
\end{equation}
These two wave equations describe eight degrees of freedom: five for a massive 
spin-$2$ particle, two for a massless spin-$2$ particle and one for a massive 
spin-$0$ particle.

The momentum space propagators of eqs. (23) and (24) are given by
\begin{equation}
D_{\mu\nu,\alpha\beta}(k) =\frac{\frac{-i}{2}\left(\eta_{\mu\alpha}
\eta_{\nu\beta} +\eta_{\mu\beta}\eta_{\nu\alpha} - \eta_{\mu\nu}
\eta_{\alpha\beta}\right)}{k^2(k^2 + \lambda^{-2})},
\label{25}
\end{equation}
and
\begin{equation}
D(k) = \frac{-i}{k^2 + \lambda^{-2}},
\label{26}
\end{equation}
respectively. These propagators have a good ultraviolet behavior, so the 
standard power counting arguments can be used. Note that we can write the 
propagator (25) as
\begin{equation}
D_{\mu\nu,\alpha\beta}(k) = \frac{\frac{-i}{2}\left(\eta_{\mu\alpha}
\eta_{\nu\beta} +\eta_{\mu\beta}\eta_{\nu\alpha} - \eta_{\mu\nu}
\eta_{\alpha\beta}\right)}{\lambda^{-2}}\left[ \frac{1}{k^2} -
\frac{1}{k^2 + \lambda^{-2}}\right].
\label{27}
\end{equation}
The minus sign between the two terms in brackets suggest the presence of a 
negative norm ghost state in massive conformal gravity. However, the theory 
might be free from ghosts if quantized correctly according to the rules of a 
conformal quantum mechanics. A similar procedure has been carried out with 
$\cal{PT}$ symmetric oscillators by using the methods of $\cal{PT}$ quantum 
mechanics \cite{Bender}. Thus a careful quantum analysis is necessary on this 
issue.


\section{The Newtonian limit}


The massive conformal gravity must be completely conformal. This means that the 
general relativistic line element  $ds^2 = g_{\mu\nu}dx^{\mu}dx^{\nu}$ must be 
replaced by the conformally invariant line element
\begin{equation}
ds^2 = \left( \varphi^{2}g_{\mu\nu} \right) dx^{\mu}dx^{\nu}.
\label{28}
\end{equation}
Accordingly, the interval $s$ between two points $P_{1}$ and $P_{2}$ along 
a parametrized timelike curve $x^{\mu} = x^{\mu}(\tau)$ is given by 
\begin{equation}
s = \int_{P_{1}}^{P_{2}} \left( \varphi^{2}g_{\mu\nu}\,\frac{dx^{\mu}}
{d\tau}\frac{dx^{\nu}}{d\tau}\right)^{\frac{1}{2}} d\tau,
\label{29}
\end{equation}
where the parameter $\tau$ is identified as the proper time. The extremization 
of the functional (29) gives the conformal geodesic equation \cite{Wood}
\begin{equation}
\frac{d^{2}x^{\lambda}}{d\tau^2} + \Gamma^{\lambda}\,\!\!_{\mu\nu}
\frac{dx^{\mu}}{d\tau}\frac{dx^{\nu}}{d\tau} +\frac{1}{\varphi}
\frac{\partial\varphi}{\partial x^{\rho}} \left( g^{\lambda\rho} + 
\frac{dx^{\lambda}}{d\tau}\frac{dx^{\rho}}{d\tau}\right) = 0,
\label{30}
\end{equation}
where
\begin{equation}
\Gamma^{\lambda}\,\!\!_{\mu\nu} = \frac{1}{2}g^{\lambda\rho}\left( 
\partial_{\mu}g_{\nu\rho} + \partial_{\nu}g_{\mu\rho} 
- \partial_{\rho}g_{\mu\nu} \right)
\label{31}
\end{equation}
is the Levi-Civita connection.

The theory presented here is independent of the gauge choice. However, 
it will be easier to find the classical results of the theory by imposing 
the unitary gauge $\varphi = \varphi_{0} =$ constant. In this case, 
the conformal geodesic equation (30) reduces to
\begin{equation}
\frac{d^{2}x^{\lambda}}{d\tau^2} + \Gamma^{\lambda}\,\!\!_{\mu\nu}
\frac{dx^{\mu}}{d\tau}\frac{dx^{\nu}}{d\tau} = 0,
\label{32}
\end{equation}
which is just the general relativistic geodesic equation. As is 
well know, the Newtonian limit of such geodesic equation gives
\begin{equation}
h_{00} = -\frac{2\phi}{c^2},
\label{33}
\end{equation}
where $\phi$ is the time-independent Newtonian potential.

If we choose $\varphi = \sqrt{2\alpha/\beta}$, for simplicity, the field 
equations (7) and (8) become
\begin{equation}
W_{\mu\nu} - \lambda^{-2} G_{\mu\nu} = 0,
\label{34}
\end{equation}
and
\begin{equation}
R = 0,
\label{35}
\end{equation}
respectively. Taking into account the weak-field approximation (12) and 
the coordinate gauge condition (21), it is not difficult to see that eqs. 
(34) and (35) lead to the wave equation (23). For a time-independent field, 
the $00$ component of this wave equation reduces to 
\begin{equation}
\left(\nabla^{2} - \lambda^{-2} \right)\nabla^{2}h_{00}  = 0,
\label{36}
\end{equation}
where $\nabla^2$ is the Laplacian operator. 

Substituting eq. (33) into eq. (36), we obtain
\begin{equation}
\left( \nabla^{2} - \lambda^{-2} \right)\nabla^{2}\phi=0.
\label{37}
\end{equation}
The solution of this equation in spherical coordinates reads 
\begin{equation}
\phi(r)= a +\frac{b}{r} + c\,\frac{e^{-r/\lambda}}{r}+d\,
\frac{e^{\,r/\lambda}}{r},
\label{38}
\end{equation}
where $a$, $b$, $c$ and $d$ are arbitrary constants. Since $\lambda > 0$, 
the last term in eq. (38) goes to infinity as $ r \longrightarrow \infty$, 
which is unphysical. In addition, at a small distance ($r \ll \lambda$) 
from a particle of mass $M$, the potential (38) must reduces to the usual 
Newtonian potential
\begin{equation}
\phi(r) = - \frac{GM}{r}.
\label{39}
\end{equation}
So we set $a=0$, $b=-GM/(1+\gamma)$, $c = - \gamma GM/(1+\gamma)$ and 
$d = 0$, and eq. (38) becomes
\begin{equation}
\phi(r) = - \frac{GM}{r(1+\gamma)}\left[ 1+\gamma e^{-r/\lambda} \right],
\label{40}
\end{equation}
where $\gamma$ determines the strength of the Yukawa potential relative to 
the Newtonian potential. 

The constant $\gamma$ and the range $\lambda$ of the Yukawa potential must be 
determined by experimental tests. The rotation curves of the major number of 
galaxies are reproduced with $\gamma  = -0.92$ and $\lambda = 20 - 30 \, 
\mbox{kpc}$, which requires that $m \sim  10^{-26} \, \mbox{eV}/\mbox{c}^2$ 
\cite{Sanders, Griv}. The maximum length scale of galaxies is in some way 
determined by $\lambda$. On scales larger than $\lambda$ the repulsive Yukawa 
potential cuts off and the attractive Newtonian potential remains, which allows 
the formation of galaxy clusters. 

It is worth noting that the gravitational potential present here is not related 
with the Newtonian solutions discussed in the work of Flanagan \cite{Flanagan} 
and later work of Mannheim \cite{Mannheim3}. The theory of conformal gravity 
with dynamical mass generation considered by the authors leads to 
\begin{equation}
R = \frac{6\rho}{\varphi_{0}^2c}
\label{41}
\end{equation} 
in vacuum, where $\rho$ is the source density.  We can readily see that this 
equation differs from eq. (35) of massive conformal gravity with the unitary 
gauge $\varphi = \varphi_{0}$ imposed.


\section{Final remarks}


The theory presented here might play an important role on both atomic and 
cosmological scales. The use of the correct conformal quantization method 
may show that the theory is renormalizable and unitary. At the same time, the 
modified potential (40) seems to be a good candidate to describe cosmological 
phenomena such as the galaxies rotation curves. These issues are under 
investigation now. The coupling of the theory with matter fields, which is 
important for a complete description of the theory, will be investigated in the 
future.


\end{document}